\documentclass[cameraready]{Interspeech}

\title{Reducing the Offline-Streaming Gap for Unified ASR Transducer with Consistency Regularization}

\author[affiliation={1}, orcid=0000-0002-8697-832X, equalcontribution]{Andrei}{Andrusenko}
\author[affiliation={1}, orcid=0009-0005-7845-5042, equalcontribution]{Vladimir}{Bataev}
\author[affiliation={1}]{Lilit}{Grigoryan}
\author[affiliation={1}]{Nune}{Tadevosyan}
\author[affiliation={2}]{Vitaly}{Lavrukhin}
\author[affiliation={2}]{Boris}{Ginsburg}

\address{
    \mbox{$^1$NVIDIA, Armenia \qquad $^2$NVIDIA, USA}
}

\email{\{aandrusenko, vbataev, lgrigoryan,ntadevosya, vlavrukhin\}@nvidia.com}


\usepackage{comment}
\usepackage{amsmath}
\usepackage{multirow}
\usepackage{hyperref}

\begin{document}

\maketitle

\begin{abstract}

Unification of automatic speech recognition (ASR) systems reduces development and maintenance costs, but training a single model to perform well in both offline and low-latency streaming settings remains challenging. We present a Unified ASR framework for Transducer (RNNT) training that supports both offline and streaming decoding within a single model, using chunk-limited attention with right context and dynamic chunked convolutions. To further close the gap between offline and streaming performance, we introduce an efficient Triton implementation of mode-consistency regularization for RNNT (MCR-RNNT), which encourages agreement across training modes. Experiments show that the proposed approach improves streaming accuracy at low latency while preserving offline performance and scaling to larger model sizes and training datasets. The proposed Unified ASR framework and the English model checkpoint are open-sourced.

\end{abstract}

\section{Introduction}

Deploying automatic speech recognition (ASR) systems commonly requires both high-accuracy offline transcription and low-latency streaming performance. Maintaining separate models for these regimes increases the cost of model development, training, validation, and deployment. All of these motivate efforts to train a single unified model~\cite{Tripathi2020TransformerTO,Yu2020DualmodeAU,Yao2021WeNetPO,Liu2022LearningAD}. 


The Transducer ASR architecture (RNNT)~\cite{graves2012rnnt} is natural for streaming inference, as the RNNT decoder depends only on the previous token output.
However, the most popular encoder model based on Conformer architecture~\cite{conformer} introduces a training-inference mismatch in its multi-head attention (MHA) and convolution blocks during chunked decoding. 

It is common to use chunk-limited attention to adapt the MHA block for streaming inference~\cite{Chen2020DevelopingRS,Moritz2020StreamingAS}. To limit context in convolutions, causal convolutions can be used, which prevent the current audio frame from accessing future context. These methods enable effective streaming adaptation even under minimal delays of 1-2 frames~\cite{Noroozi2023StatefulCW}. However, all of these lead to a noticeable accuracy degradation compared to offline models.

Several complementary approaches have demonstrated the importance of providing the right context for streaming models, including unified training. A Zipformer-based unified framework~\cite{Sharma2025UnifyingSA} trains with chunk-limited attention and dynamic right-context, reporting that increasing right-context closes much of the quality gap. Time-Shifted Contextual Attention (TSCA) and Dynamic Right Context (DRC) masking~\cite{Le2024ImprovingSS} likewise incorporate future information in a chunked streaming pipeline and report consistent improvement on LibriSpeech by better leveraging in-context future information. Dynamic Chunk Convolution (DCConv)~\cite{Li2023DynamicCC} replaces causal convolutions with a chunk-aware alternative to ensure training better mimics streaming inference by preventing models from seeing future frames across chunk boundaries. Orthogonally, All-in-One ASR~\cite{Moriya2025AllinOneAU} unifies not only offline/streaming but also multiple ASR paradigms (CTC/AED/Transducer) via a multi-mode joiner, emphasizing broader model footprint reduction.


Despite the progress in unified ASR modeling, low-latency streaming (e.g., a look-ahead latency budget of less than 0.5s) remains challenging. The smaller the right context, the stronger the mode conflict becomes, and unified models often exhibit a sharp degradation in offline or streaming regimes. Moreover, while prior work reports strong, unified results on limited training datasets. The interaction between unification mechanisms and large-scale training remains underexplored, hindering stable, low-latency performance across diverse domains.

In this work, we present a Unified ASR framework for a Transducer modeling that combines chunk-limited attention with right context and dynamic chunked convolutions to adapt the model to both decoding modes. We further introduce a mode-consistency regularization for RNNT (MCR-RNNT), implemented efficiently with GPU kernels, to explicitly reduce the gap between offline and streaming behaviors within a single set of parameters. We show that the proposed method improves model performance in low-latency streaming while preserving offline quality and remains effective at scale, reaching 5.76\% AVG WER on the open ASR Leaderboard\cite{srivastav2025openasrleaderboardreproducible} for English.

The key contribution of our work can be formulated as:
\begin{itemize}
    \item A Unified ASR Transducer framework\footnote{\scriptsize{Will be released soon}} that supports offline and streaming modes with shared parameters.
    \item An efficient GPU implementation of mode-consistency regularization loss for the RNNT model.
    \item State-of-the-art Unified RNNT model\footnote{\scriptsize{\url{https://huggingface.co/nvidia/parakeet-unified-en-0.6b}}} on the English language with punctuation and capitalization support.
    \end{itemize}


\section{Method}

We train a single RNNT model with shared parameters to support both offline and streaming decoding. The model follows the standard Transducer design with encoder, predictor, and joint. Our encoder uses Conformer-style blocks with multi-head attention (MHA) and convolution modules. To enable streaming, we restrict MHA and convolution context during training.

\subsection{Chunk-limited attention with right context}

For streaming training, MHA is constrained by a chunked mask with three parts: left context (L), current processing chunk (C), and right context (R). At each step, frames of the encoder layer in the current chunk may attend to all frames from the left context L, frames within the current chunk C, and up to R future frames beyond the current chunk boundary. To support multiple latency targets with a single model, we sample diverse C and R values from a predefined set during training.

\subsection{Dynamic chunk convolution}

Standard convolutions may depend on future frames that are unavailable in streaming, while causal convolutions lose useful future context and often reduce accuracy. Dynamic Chunk Convolution (DCConv) addresses this by making convolutions chunk-aware and better matched to streaming inference~\cite{Li2023DynamicCC}.

Following this idea, we use DCConv in each Conformer block. In streaming mode, before the depthwise convolution, we reshape the hidden states into chunks according to the current chunk size C, together with left and right contexts equal to $\frac{\mathrm{kernel\_size}-1}{2}$, while sharing the same convolution parameters across offline mode. This reduces the train-inference mismatch introduced by chunked streaming decoding.

\subsection{Unified training strategy}

We consider two main strategies for unified model training:

\textbf{Single-mode (SM)}. Each optimization step samples a mode type $m \in \{\text{offline}, \text{streaming}\}$ with probability $p_{\text{off}} $ and runs one forward-backward pass in that mode:

\begin{equation}
    \mathcal{L}_{\text{SM}} = \mathcal{L}_{\text{RNNT}}^{(m)}.
\end{equation} 

\textbf{Dual-mode (DM)}. Each step runs both modes on the same input batch and combines their RNNT losses:
\begin{equation}
    \mathcal{L}_{\text{DM}} = \alpha \, \mathcal{L}_{\text{RNNT}}^{\text{off}}
    + (1-\alpha)\, \mathcal{L}_{\text{RNNT}}^{\text{str}},
\end{equation} 
where $\alpha \in[0,1]$ represents the offline mode weight (by analogy with $p_{\text{off}}$ in SM). This approach doubles the computational resources per training step compared to SM, but more directly couples the two modes during optimization. 

\begin{figure}[t!]
  \centering
  \includegraphics[width=7.5cm]{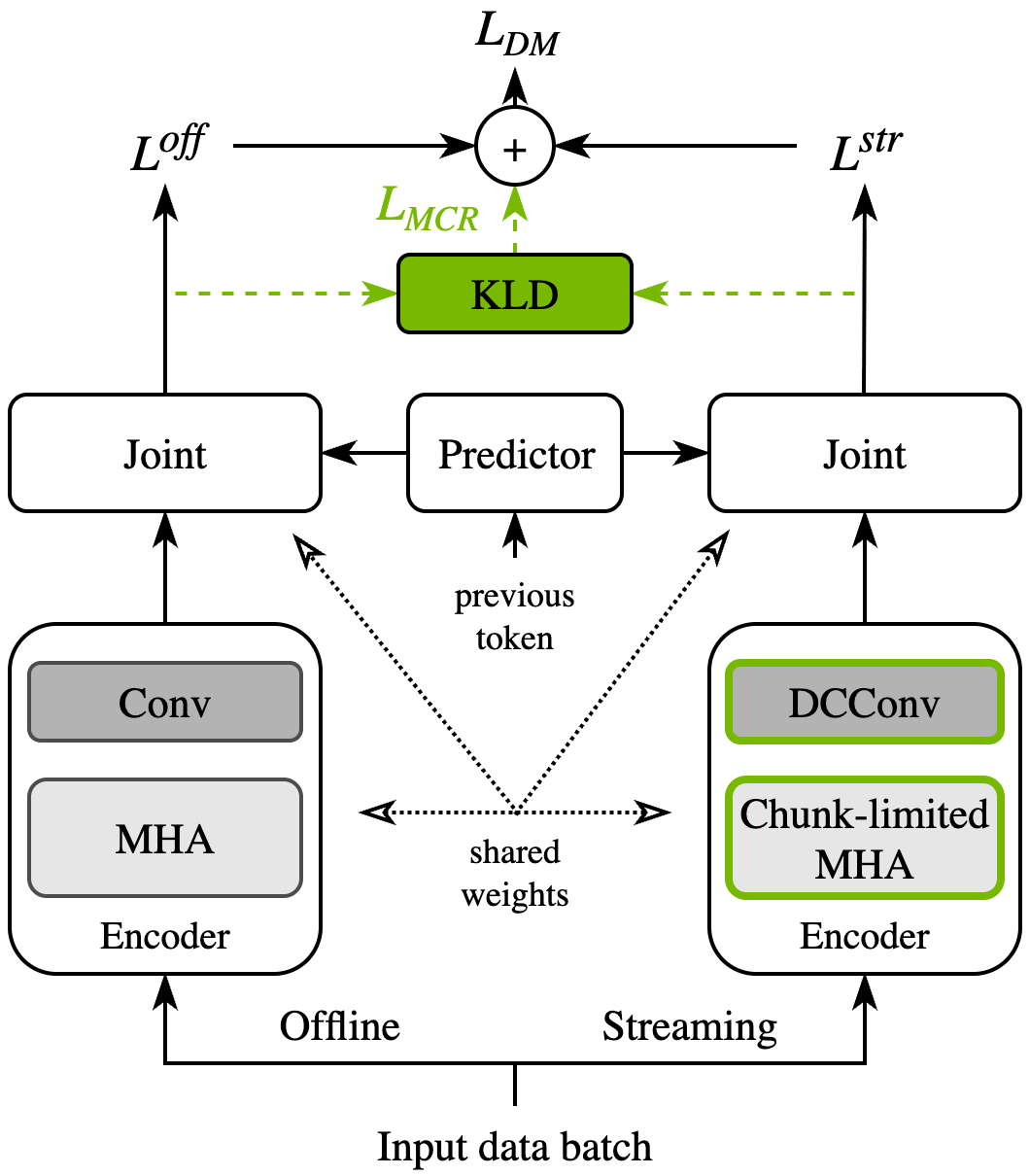}
  \caption{Unified Transducer training in dual mode with mode-consistency regularization (MCR-RNNT) loss.}
  \label{fig:unified-asr}
\end{figure}

\begin{table*}[t!]
  \centering
  \caption{Comparison of Average WER (\%) on Open ASR Leaderboard for offline and streaming inference mode with various latency constraints. Left context was set to 5.6s (70 frames). Latency is defined as the sum of the chunk and the right context size: 2.08s=1.04+1.04, 1.12s=0.56+0.56, 0.56s=0.16+0.40, 0.40s=0.08+0.32, 0.32s=0.08+0.24, 0.24s=0.08+0.16, and 0.16s=0.08+0.08.}
  \label{tab:wer_results}
  \resizebox{\textwidth}{!}{%
  \begin{tabular}{l |c| c c c c c c | c}
    \toprule
    \multirow{2}{*}{\textbf{Model Setup}} & \multirow{2}{*}{\textbf{Offline}} & \multicolumn{7}{c}{\textbf{Streaming Latency (seconds)}} \\
    \cmidrule(lr){3-9}
    & & \textbf{2.08s} & \textbf{1.12s} & \textbf{0.56s} & \textbf{0.40s} & \textbf{0.32s} & \textbf{0.24s} & \textbf{0.16s} \\
    \midrule
    \textit{L-size models ($\sim$128M params) -- 120k hours of norm data} \\
    \addlinespace[0.5ex]
    Baseline (Offline)           & \textbf{6.47} & 6.92 & 8.21 & 13.56 & 26.51 & 49.46 & 78.67 & 94.05 \\
    Baseline (Streaming)      & 7.75 & \textit{8.39} & 8.02 & 8.36 & \textit{11.47} & 9.44  & \textit{10.01} & \textbf{9.84} \\
    Mamba2 + DCConv (Streaming)      & 7.89 & 8.41 & 8.62 & 8.68 & 8.66 & 8.95 & 9.53 & 10.52 \\
    \midrule
    Unified single-mode (SM)                  & 6.66 & 7.71 & 7.46 & 7.98 & 9.40 & 10.96 & 13.33 & 17.16 \\
    Unified dual-mode (DM)                  & 6.69 & 7.14 & 7.48 & 8.12 & 9.86 & 12.48 & 16.91 & 22.45 \\
    \textbf{Unified DM + MCR-RNNT (Ours)}        & 6.63 & \textbf{6.86} & \textbf{7.09} & \textbf{7.47} & \textbf{7.83} & \textbf{8.24} & \textbf{9.04} & 10.51 \\
    \midrule
    \midrule
    \textit{XL-size models ($\sim$600M params) -- 240k hours of PC data} \\
    \addlinespace[0.5ex]
    Parakeet-TDT-0.6b-v2~\cite{model:tdt-v2}        & 6.04 & 7.99 & 22.83 & 69.55 & 95.12 & 97.32 & 99.10 & 99.47 \\
    Nemotron-Speech-Streaming-En-0.6b~\cite{model:nemotrom-streaming}     & 7.05 & \textit{7.51} & 7.08 & 7.22 & \textit{10.91} & 7.78  & \textit{8.18} & \textbf{7.92} \\
    \midrule
    \textbf{(1) Unified DM + MCR-RNNT 0.6B (Ours, larger right cont.)} & \textbf{5.76} & \textbf{5.97} & \textbf{6.14} & \textbf{6.44} & 6.96 & 7.72 & 9.51 & 12.73\\
    \textbf{(2) Unified DM + MCR-RNNT 0.6B (Ours, balanced)} & 5.91 & 6.14 & 6.29 & 6.52 & \textbf{6.70} & \textbf{6.92} & \textbf{7.35} & 8.44 \\
    \bottomrule
  \end{tabular}%
  }
\end{table*}

\subsection{Consistency regularization for RNNT}

Unified masking and chunked convolutions improve mode robustness, but low-latency streaming (less than 0.5s) can still degrade sharply because the same parameters must represent two regimes with different available context. We address this by adding a mode-consistency regularization term, inspired by prior works on consistency regularization.

As a first step, we extended CR-CTC~\cite{Yao2024CRCTCCR} to unified ASR by applying symmetric consistency between offline and streaming CTC posteriors in hybrid CTC-RNNT training. However, this consistently degraded streaming RNNT accuracy, even though offline performance remained strong. We attribute this to an objective mismatch: the auxiliary CTC loss encourages frame-synchronous, locally confident token predictions, while low-latency streaming RNNT requires richer encoder representations under a limited future context. As a result, the shared encoder is biased toward alignments that are easier to realize offline but less suitable for streaming Transducer decoding.

TCR~\cite{Tseng2024TransducerCR} work applies consistency to pruned RNNT outputs for augmented views and relies on occupation-based weighting to handle the large alignment space. In our work, we target mode consistency between offline and streaming decoding and require a practical full-lattice formulation for unified training. This setup differs, and the alignment can vary significantly between modes due to the greater flexibility of offline representations. Additionally, no publicly available implementation was available. Therefore, we developed our own mode-consistency regularization loss for RNNT (MCR-RNNT).

Our implementation uses full RNNT joint logits \(z^{(\widetilde{t})}, z^{(\widetilde{s})} \in \mathbb{R}^{T \times (U+1) \times V}\) from teacher and student modes, respectively. It computes KL divergence (KLD) directly from raw logits inside a fused GPU kernel integrated with PyTorch~\cite{paszke2019pytorch} autograd. We chose Triton~\cite{tillet2019triton} for the implementation because it is easier to maintain, delivers near-CUDA performance, and is highly portable.
At each \((t,u)\), with $p=\mathrm{softmax}(z^{(\widetilde{t})}_{t,u,:})$ and $q=\mathrm{softmax}(z^{(\widetilde{s})}_{t,u,:})$, we compute
\begin{equation}
\begin{split}
\mathcal{L}_{MCR} = \mathcal{L}_{KL}(p \mid q) = \sum_{v=1}^{V} p_v\left(\log p_v-\log q_v\right)
\end{split}
\label{eq:mcr_loss}
\end{equation}

For symmetric consistency regularization, we use
\begin{equation}
\begin{split}
\mathcal{L}_{MCR-Sym} = \frac{1}{2}[\mathcal{L}_{KL}(p \mid q) + \mathcal{L}_{KL}(q \mid p)] \\ 
= \frac{1}{2}\sum_{v=1}^{V}(p_v-q_v)(\log p_v-\log q_v)
\end{split}
\label{eq:mcr_synloss}
\end{equation}

We reduce per-utterance losses by normalizing consistency over valid $(t,u)$ lattice elements.

Since explicit materialization of $\log\mathrm{softmax}$ for the full $[T,U+1,V]$ joint tensor (including extra batch dimension) is prohibitive for large vocabulary size, we compute log-softmax and KLD on-the-fly, recomputing them in backward pass to produce the gradient. This design mirrors the RNNT loss memory strategy in NeMo~\cite{kuchaiev2019nemo}, and imposes nearly zero memory overhead and tiny computational overhead compared to RNNT loss. 

Along with standard KL loss with full RNNT Joint output, we also investigated different consistency regularization strategies: computing KLD over $\{p_{\text{blank}}, p_{\text{target}}, 1 - p_{\text{blank}} - p_{\text{target}}\}$ probability distribution, and also a separate variant of using lattice posteriors instead of output probabilities. Early experiments showed that the original KLD over the full joint output performs better and is more stable across these variants.

Our final objective function in dual-mode training is
\begin{equation}
\mathcal{L}_{DM} =
\alpha \, \mathcal{L}_{\text{RNNT}}^{\text{off}}
+ (1-\alpha)\, \mathcal{L}_{\text{RNNT}}^{\text{str}}
+ \lambda \, \mathcal{L}_{\text{MCR}},
\label{eq:full_objective}
\end{equation}
where $\lambda \ge 0$ controls the strength of MCR-RNNT loss.

\section{Experimental setup}

\subsection{ASR modeling and evaluation}

As the main ASR architecture, we used RNNT model based on FastConformer encoder~\cite{rekesh2023fastconformer} with 123M parameters. The input features are 128-dim FBanks with x8 initial subsampling. The prediction network (decoder) is a single-layer LSTM with 640 units, which increased the total model size to 128M parameters. All models were trained for 100K steps using a cosine annealing LR scheduler. The maximum LR was 1e-3 with 15K LR warmup. The training was done in the NeMo~\cite{kuchaiev2019nemo} framework with 32 NVIDIA A100 GPUs using dynamic bucketing~\cite{elasko2024EMMeTTEM}.

For model training, we used a subset of $\sim$120,000 hours of labeled English speech (with normalized transcripts) from the public Granary dataset~\cite{Koluguri2025GranarySR}. For text tokenization, we used BPE tokenizer~\cite{bpe} with 1024 tokens.

To evaluate the ASR results, we used the Open ASR Leaderboard for English~\cite {srivastav2025openasrleaderboardreproducible}. For all models, we computed average WER across 8 different test sets, including AMI, Earnings22, Gigaspeech, Librispeech, SPGI, TEDLIUM, and VoxPopuli open test sets. We believe that testing models across such a wide range of data domains yields more robust results. 

Inference evaluation was performed in an efficient greedy decoding mode~\cite{bataev2024labellooping,galvez24_speedoflight} with batch size 128. During offline decoding, the model had access to the whole input audio file. In streaming, we used stateful chunk-based decoding with fixed parameters of L, C, and R with a step size of C, discarding encoder representations for L and R context at each step. We define the overall theoretical worst-case latency as $C+R$. 

\subsection{Streaming baseline}

As a pure streaming baseline, we trained a cache-aware streaming model~\cite{Noroozi2023StatefulCW} with the same RNNT model parameters. For chunk-limited attention masks, we used a default multi-look-ahead setup from the original paper [[70], [13,6,1,0]], where the first value is the left context and the second set of values is the dynamic look-ahead size sampled uniformly. This model has no right context for the attention mask by design. 

We also implemented a streaming version of the FastConformer encoder for RNNT, replacing the MHA with the Mamba2~\cite{Dao2024TransformersAS} block. As a convolutional block in this model, we used our DCConv implementation with chunk sizes sampled from [1,2,7,13]. The probability of switching between shared full-context convolutions and DCConv was set to 0.5. 

\subsection{Unified ASR setup}

For all the streaming modes in unified training (SM and DM), we used the same encoder parameters. Context size for chunk-limited attention mask [L, C, R] was sampled from a predefined list [[70],[1,2,7,13],[0,1,2,3,5,7,13,26]] represented in frames after initial x8 subsampling (1 frame here is equal to 80ms) at each training step. These parameters demonstrated the best results during the initial experiments with parameter search. 

In the unified dual-mode training, we reduce the batch size twice to match the computational complexity of the baselines and single-mode training.

\subsection{Data and model size scaling}

In addition to all experiments, we trained the best-performing unified setup using an RNNT XL-size model ($\sim$600M parameters) on 280,000 hours of English data from the Granary dataset, including punctuation and capitalization (PC). We train XL models for 300K steps with an LR of 5e-4 and cosine annealing.

\section{Results}

\begin{table}[t!]
  \centering
  \caption{Average WER (\%) on Open ASR Leaderboard for different training configurations of KLD teacher, KLD weight $\lambda$, and offline $\alpha$ for the same unified RNNT L-size model.}
  \label{tab:ablation_results_joined}
  \resizebox{\columnwidth}{!}{
  \begin{tabular}{lcccc}
    \toprule
    \textbf{Configuration Variable} & \textbf{Offline} & \textbf{1.12s} & \textbf{0.56s} & \textbf{0.32s} \\
    \midrule
    \textit{KLD Teacher ($\lambda$=0.1, $\alpha$=0.5)} \\
    \addlinespace[0.5ex]
    Offline  & 6.64 & 7.33 & 8.55 & 15.86 \\
    Streaming & 6.82 & 7.67 & 7.85 & 8.56  \\
    Symmetric           & 6.61 & 7.16 & 7.60 & 8.34  \\
    \midrule
    \textit{KLD weight $\lambda$ (sym, $\alpha$=0.5)} \\
    \addlinespace[0.5ex]
    0.1 & 6.61 & 7.16 & 7.60 & 8.34 \\
    0.2 & 6.66 & 7.15 & 7.50 & 8.17 \\
    0.3 & 6.63 & 7.09 & 7.47 & 8.24 \\
    0.5 & 6.71 & 7.18 & 7.60 & 8.50 \\
    \midrule
    \textit{Offline weight $\alpha$ (sym, $\lambda$=0.3)} \\
    \addlinespace[0.5ex]
    0.3 & 6.68 & 7.16 & 7.49 & 8.11 \\
    0.5 & 6.63 & 7.09 & 7.47 & 8.24 \\
    0.7 & 6.61 & 7.17 & 7.62 & 8.72 \\
    \bottomrule
  \end{tabular}
  }
\end{table}

Table~\ref{tab:wer_results} presents the main evaluation results for the considered models in the offline and streaming decoding scenarios. 

\subsection{L-size models with 120K hours of normalized data}

The offline baseline model demonstrated the best offline decoding accuracy. However, the model degraded substantially during streaming inference after reducing the latency to less than 1.12s. The streaming baseline model showed high robustness even at a 0.16s latency, but it did not gain much improvement from having the entire audio during decoding (offline mode), significantly underperforming the offline baseline due to a lack of contextual capabilities (chunk-limited attention and causal convolutions). The model also struggles with latency values (italic font) that were not included in the look-ahead list during training. Replacing MHA with the Mamba2 block and DCConv yielded comparable results to the streaming baseline.

Next, we obtained results for a standard unified training in both single and dual-mode settings. Despite identical training parameters and computational resources, single-mode outperformed dual-mode in offline and streaming decoding scenarios. However, Unified SM began to degrade noticeably when the latency was reduced to less than 0.56s-0.40s, making the streaming baseline more appropriate for low-latency streaming.  

The use of the proposed MCR-RNNT loss during unified dual-mode training demonstrated superior model performance in offline and streaming decoding (up to 0.24s latency), outperforming all the considered models. The model is only slightly inferior to the streaming baseline at 0.16s latency. We suppose that the MCR-RNNT loss explicitly reduces the representation gap between offline and streaming modes at the Transducer output level. As a result, the shared model is encouraged to learn predictions that remain stable under different context constraints, leading to a better trade-off (Pareto frontier) between offline accuracy and low-latency streaming performance. 

\subsection{XL-size models with 280K hours of PC data}

The scaling experiments showed a complementary improvement for the proposed Unified RNNT method. We trained two models with different sets of right context parameters, balancing offline and streaming performance. The first model (1) with a larger right context outperforms the strong open-source Nemotron-Streaming-0.6b~\cite{model:nemotrom-streaming} model in offline and streaming, up to 0.32s latency. The model (1) showed even better results in offline than Parakeet-TDT-0.6b-v2~\cite{model:tdt-v2} trained on the same Granary dataset. Obtained 5.76\% AVG WER in offline almost reaches the best results 5.63\% from Canary-Qwen-2.5B~\cite{model:canary-qwen} (pure offline model), making our model SOTA Unified RNNT. 

The second (2) Unified RNNT model (trained with smaller right context values) demonstrated the trade-off results between offline and streaming decoding scenarios, still outperforming strong offline and streaming models from the Open ASR Leaderboard. The proposed model (2) is only slightly inferior to Nemotron-Streaming-0.6b at 0.16s latency point.

\subsection{Ablation studies}

In table~\ref{tab:ablation_results_joined}, we investigated different configurations of KLD type, KLD weight $\lambda$, and offline weight $\alpha$ for the proposed Unified training. The results showed that a symmetric KLD loss with $\lambda=0.3$ yields the best trade-off between offline and streaming model performance. Changing the offline weight $\alpha$ can slightly shift the balance towards improving offline or streaming results. We recommend using $\alpha=0.5$ as a starting point here. 

\begin{figure}[t]
  \centering
  \includegraphics[width=\columnwidth]{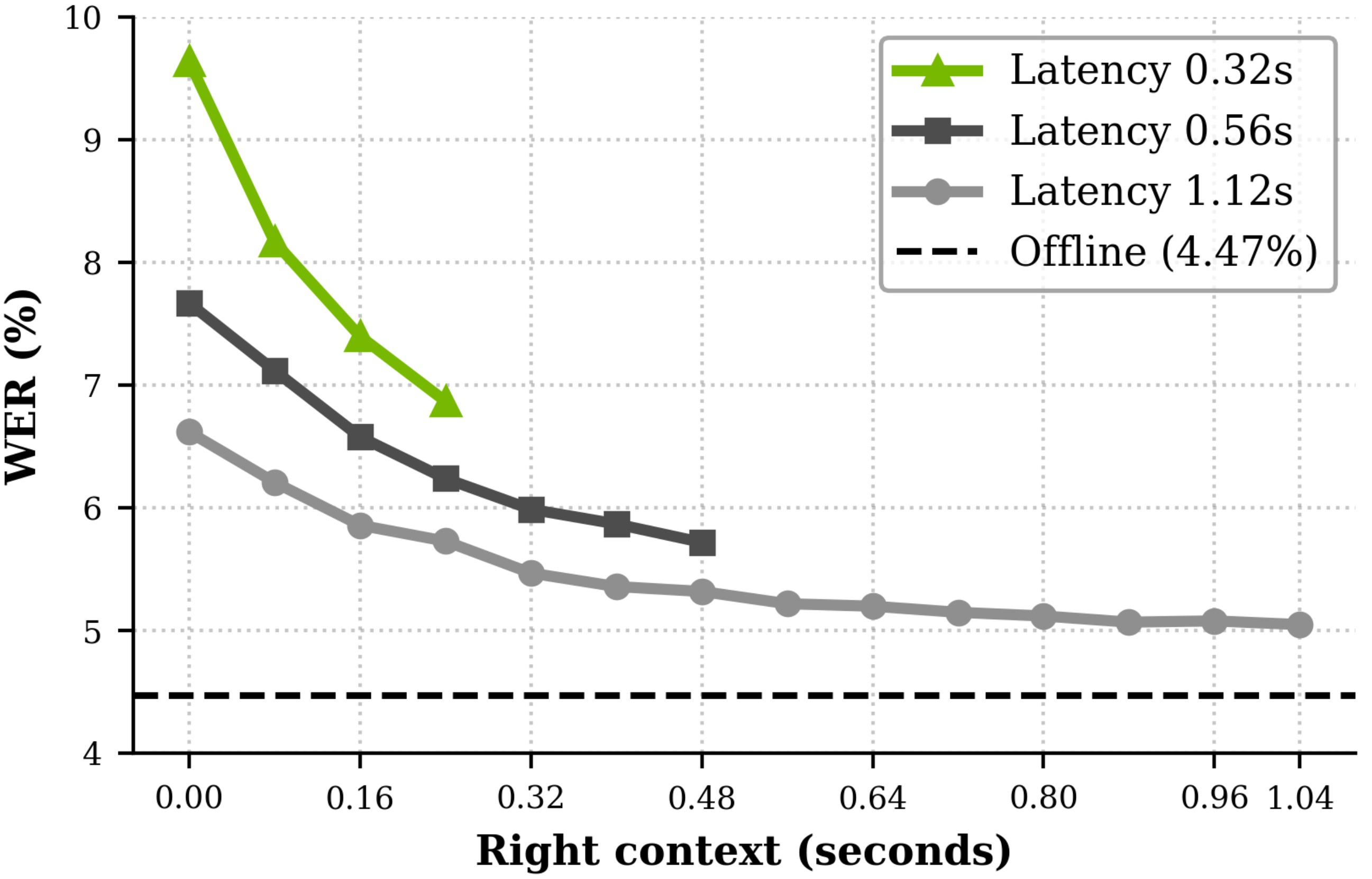}
  \caption{LibrisSpeech test other WER (\%) for different chunk and right context balance during inference under fixed total latency budgets from 0.32s to 1.12s (chunk + right context).}
  \label{fig:wer_rc_plot}
\end{figure}

Figure~\ref{fig:wer_rc_plot} demonstrates the dependency of streaming decoding accuracy on using more right context under a fixed latency budget for the same unified model. The larger right context enables lower WER, especially in a low-latency setup. However, reducing the chunk size increases the model decoding time. 

As future work, we will implement a cache-passing mechanism to enable efficient streaming decoding for the proposed unified RNNT models. Currently, we recalculate the left context at each chunk step C, which slows inference speed.  

\section{Conclusion}

We propose a new Unified ASR framework that achieves robust Transducer performance in both offline and streaming decoding scenarios. In addition to using chunk-limited attention and dynamic chunked convolutions, we introduce a novel mode-consistency regularization loss (MCR-RNNT), which further reduces the gap between offline and streaming encoder behavior. The proposed method demonstrated superior performance compared to the standard unified training methods, maintaining its advantages even for data and model size scaling. As a result, we obtained a SOTA Unified RNNT model with punctuation and capitalization support that outperforms strong open-source baselines in both offline and streaming scenarios (with up to 0.24s latency). The proposed Unified framework and English model checkpoint are open-sourced.

\bibliographystyle{IEEEtran}
\bibliography{mybib}

@article{conformer,
  author={Anmol Gulati and James Qin and Chung-Cheng Chiu and others},
  title={Conformer: Convolution-augmented Transformer for Speech Recognition},
  year={2020},
  journal={Proc. Interspeech 2020},
  pages={5036--5040},
  doi={10.21437/Interspeech.2020-3015}
}

@article{Tripathi2020TransformerTO,
  title={Transformer Transducer: One Model Unifying Streaming and Non-streaming Speech Recognition},
  author={Anshuman Tripathi and Jaeyoung Kim and Qian Zhang and Han Lu and Hasim Sak},
  journal={ArXiv},
  year={2020},
  volume={abs/2010.03192},
}

@article{Yu2020DualmodeAU,
  title={Dual-mode ASR: Unify and Improve Streaming ASR with Full-context Modeling},
  author={Jiahui Yu and Wei Han and Anmol Gulati and Chung-Cheng Chiu and Bo Li and Tara N. Sainath and Yonghui Wu and Ruoming Pang},
  journal={ICLR},
  year={2021},
}

@inproceedings{Yao2021WeNetPO,
  title={WeNet: Production Oriented Streaming and Non-Streaming End-to-End Speech Recognition Toolkit},
  author={Zhuoyuan Yao and Di Wu and Xiong Wang and Binbin Zhang and Fan Yu and Chao Yang and Zhendong Peng and Xiaoyu Chen and Lei Xie and Xin Lei},
  booktitle={Interspeech},
  year={2021},
}

@article{Liu2022LearningAD,
  title={Learning a Dual-Mode Speech Recognition Model VIA Self-Pruning},
  author={Chunxi Liu and Yuan Shangguan and Haichuan Yang and Yangyang Shi and Raghuraman Krishnamoorthi and Ozlem Kalinli},
  journal={SLT},
  year={2022},
  pages={273-279},
}

@inproceedings{graves2012rnnt,
  title={Sequence transduction with recurrent neural networks},
  author={Graves, Alex},
  booktitle={ICML},
  year={2012}
}

@article{Chen2020DevelopingRS,
  title={Developing Real-Time Streaming Transformer Transducer for Speech Recognition on Large-Scale Dataset},
  author={Xie Chen and Yu Wu and Zhenghao Wang and Shujie Liu and Jinyu Li},
  journal={ICASSP},
  year={2020},
  pages={5904-5908},
}

@article{Moritz2020StreamingAS,
  title={Streaming Automatic Speech Recognition with the Transformer Model},
  author={Niko Moritz and Takaaki Hori and Jonathan Le Roux},
  journal={ICASSP},
  year={2020},
  pages={6074-6078},
}

@article{Noroozi2023StatefulCW,
  title={Stateful Conformer with Cache-Based Inference for Streaming Automatic Speech Recognition},
  author={Vahid Noroozi and Somshubra Majumdar and Ankur Kumar and Jagadeesh Balam and Boris Ginsburg},
  journal={ICASSP},
  year={2023},
}

@article{Sharma2025UnifyingSA,
  title={Unifying Streaming and Non-streaming Zipformer-based ASR},
  author={Bidisha Sharma and Karthik Pandia Durai and Shankar Venkatesan and Jeena Prakash and Shashi Kumar and Malolan Chetlur and Andreas Stolcke},
  journal={ACL},
  year={2025},
}

@article{Le2024ImprovingSS,
  title={Improving Streaming Speech Recognition With Time-Shifted Contextual Attention And Dynamic Right Context Masking},
  author={Khanh Le and Duc Thanh Chau},
  journal={Interspeech},
  year={2024},
}

@article{Li2023DynamicCC,
  title={Dynamic Chunk Convolution for Unified Streaming and Non-Streaming Conformer ASR},
  author={Xilai Li and Goeric Huybrechts and S. Ronanki and Jeffrey J. Farris and S. Bodapati},
  journal={ICASSP},
  year={2023},
}

@inproceedings{Moriya2025AllinOneAU,
  title={All-in-One ASR: Unifying Encoder-Decoder Models of CTC, Attention, and Transducer in Dual-Mode ASR},
  author={Takafumi Moriya and Masato Mimura and Tomohiro Tanaka and Hiroshi Sato and Ryo Masumura and Atsunori Ogawa},
  journal={ASRU},
  year={2025},
}

@article{Yao2024CRCTCCR,
  title={CR-CTC: Consistency regularization on CTC for improved speech recognition},
  author={Zengwei Yao and Wei Kang and Xiaoyu Yang and Fangjun Kuang and Liyong Guo and Han Zhu and Zengrui Jin and Zhaoqing Li and Long Lin and Daniel Povey},
  journal={ICLR},
  year={2025},
}

@article{Tseng2024TransducerCR,
  title={Transducer Consistency Regularization For Speech to Text Applications},
  author={Cindy Tseng and Yun Tang and Vijendra Raj Apsingekar},
  journal={SLT},
  year={2024},
}

@inproceedings{bpe,
    title = "Neural Machine Translation of Rare Words with Subword Units",
    author = "Sennrich, Rico  and Haddow, Barry and Birch, Alexandra",
    booktitle = "Proceedings of the 54th Annual Meeting of the Association for Computational Linguistics",
    year = "2016",
}

@inproceedings{rekesh2023fastconformer,
  title={{Fast Conformer} with linearly scalable attention for efficient speech recognition},
  author={Rekesh, Dima and Koluguri, Nithin Rao and Kriman, Samuel and  and others},
  booktitle={Automatic Speech Recognition and Understanding Workshop (ASRU)},
  year={2023},
}

@misc{srivastav2025openasrleaderboardreproducible,
      title={Open ASR Leaderboard: Towards Reproducible and Transparent Multilingual and Long-Form Speech Recognition Evaluation}, 
      author={Vaibhav Srivastav and Steven Zheng and Eric Bezzam and Eustache Le Bihan and Nithin Koluguri and Piotr Żelasko and Somshubra Majumdar and Adel Moumen and Sanchit Gandhi},
      year={2025},
      eprint={2510.06961},
      archivePrefix={arXiv},
      primaryClass={cs.CL},
}

@inproceedings{tillet2019triton,
  title={Triton: an intermediate language and compiler for tiled neural network computations},
  author={Tillet, Philippe and Kung, Hsiang-Tsung and Cox, David},
  booktitle={Proceedings of the 3rd ACM SIGPLAN International Workshop on Machine Learning and Programming Languages},
  pages={10--19},
  year={2019}
}

@inproceedings{paszke2019pytorch,
  title={{PyTorch}: An imperative style, high-performance deep learning library},
  author={Paszke, Adam and Gross, Sam and Massa, Francisco and Lerer, Adam and Bradbury, James and Chanan, Gregory and Killeen, Trevor and Lin, Zeming and Gimelshein, Natalia and Antiga, Luca and others},
  booktitle={NeurIPS},
  volume={32},
  year={2019}
}

@inproceedings{kuchaiev2019nemo,
  title={Nemo: a toolkit for building ai applications using neural modules},
  author={Kuchaiev, Oleksii and Li, Jason and Nguyen, Huyen and others},
  booktitle={{NeurIPS Workshop on Systems for ML}},
  year={2019}
}

@misc{model:tdt-v2,
  author    = {NVIDIA},
  title     = {Parakeet TDT 0.6B V2 (En)},
  year      = {2025},
  url       = {https://huggingface.co/nvidia/parakeet-tdt-0.6b-v2},
}

@misc{model:nemotrom-streaming,
  author    = {NVIDIA},
  title     = {Nemotron-Speech-Streaming-En-0.6b},
  year      = {January 2026},
  url       = {https://huggingface.co/nvidia/nemotron-speech-streaming-en-0.6b/tree/nemotron-speech-streaming-jan2026},
}

@misc{model:canary-qwen,
  author    = {NVIDIA},
  title     = {Canary-Qwen-2.5B},
  year      = {2025},
  url       = {https://huggingface.co/nvidia/canary-qwen-2.5b},
}

@article{Koluguri2025GranarySR,
  title={Granary: Speech Recognition and Translation Dataset in 25 European Languages},
  author={Nithin Rao Koluguri and Monica Sekoyan and George Zelenfroynd and Sasha Meister and Shuoyang Ding and Sofia Kostandian and He Huang and Nikolay Karpov and Jagadeesh Balam and Vitaly Lavrukhin and Yifan Peng and Sara Papi and Marco Gaido and Alessio Brutti and Boris Ginsburg},
  journal={Interspeech},
  year={2025},
  volume={abs/2505.13404},
}

@article{Dao2024TransformersAS,
  title={Transformers are SSMs: Generalized Models and Efficient Algorithms Through Structured State Space Duality},
  author={Tri Dao and Albert Gu},
  journal={ICML},
  year={2024},
}

@inproceedings{bataev2024labellooping,
  author={Bataev, Vladimir and Xu, Hainan and Galvez, Daniel and Lavrukhin, Vitaly and Ginsburg, Boris},
  booktitle={2024 IEEE Spoken Language Technology Workshop (SLT)}, 
  title={Label-Looping: Highly Efficient Decoding For Transducers}, 
  year={2024},
  pages={7-13},
}

@inproceedings{galvez24_speedoflight,
  title     = {Speed of Light Exact Greedy Decoding for RNN-T Speech Recognition Models on GPU},
  author    = {Daniel Galvez and Vladimir Bataev and Hainan Xu and Tim Kaldewey},
  year      = {2024},
  booktitle = {Interspeech 2024},
  pages     = {277--281},
  doi       = {10.21437/Interspeech.2024-1591},
}

@article{elasko2024EMMeTTEM,
  title={EMMeTT: Efficient Multimodal Machine Translation Training},
  author={Piotr Żelasko and Zhehuai Chen and Mengru Wang and Daniel Galvez and Oleksii Hrinchuk and Shuoyang Ding and Ke Hu and Jagadeesh Balam and Vitaly Lavrukhin and Boris Ginsburg},
  journal={ICASSP},
  year={2024},
}

\end{document}